# MEASURED LIGHTCURVES AND ROTATIONAL PERIODS OF 3122 FLORENCE, 3830 TRELLEBORG, AND (131077) 2000 YH105

Natasha S. Abrams, Allyson Bieryla, Sebastian Gomez, Jane Huang, John A. Lewis, Lehman H. Garrison, and Theron Carmichael
Center for Astrophysics | Harvard & Smithsonian
60 Garden St.
Cambridge MA, 02138
nsabrams@college.harvard.edu

We determined the rotational periods of 3122 Florence, 3830 Trelleborg, and (131077) 2000 YH105 with the Harvard Clay Telescope and KeplerCam at the Fred L. Whipple Observatory. We found the rotational periods to be $2.3580 \pm 0.0015$ h, $17.059 \pm 0.017$ h, and $1.813 \pm 0.00003$ h, respectively. Our measurement of 3122 Florence's period agrees with Warner (2016), who reported $2.3580 \pm 0.0002$ h.

Photometric observations of 3122 Florence, 3830 Trelleborg, and (131077) 2000 YH105 were collected using a combination of data from the Clay Telescope at Harvard University in Cambridge, MA, and KeplerCam at the Fred L. Whipple Observatory in Arizona. The Clay Telescope is a 0.4-m telescope with 13x13 arcmin FOV and an Apogee Alta U47 imaging CCD. The data collected with the Clay Telescope were in R-band Bessel-system filter. KeplerCam is a 1.2-m telescope with a 23x23 arcmin FOV; the data were taken in the Sloan i-band.

3122 Florence was chosen since it has a known rotational period of $2.3580 \pm 0.0002$ h (Warner, 2016). We chose to begin with an asteroid with a known rotational period in order to validate our methodology. Using *Minorplanet.info* to look up the brightest targets without known rotational periods, we selected 3830 Trelleborg and 2000 YH105 because they were observable by the Clay telescope for at least several months.

In our analysis, we reduced the images from the Clay Telescope using *MaximDL* (Diffraction Limited, 1997) and from KeplerCam using standard *IDL* procedures. We determined an astrometric solution using *astrometry.net* and produced a lightcurve using the *AstroImageJ* multi-aperture photometry function (Collins et al., 2017). We used the Python package *gatspy* to calculate the Lomb-Scargle periodogram and determine the rotational period (VanderPlas and Ivezić, 2015). Uncertainties in the rotational periods were determined by establishing a maximum and minimum rotational period at which the phased lightcurve no longer had recognizable rotational modulation. Amplitudes were determined by subtracting the maximum from the minimum normalized magnitude, and errors were found by measuring the scatter of one of the peaks. For normalized flux, the amplitude was found by |-2.5log(max flux/min flux)|.

3122 Florence was discovered 1981 March 2 at Siding Spring Observatory in Australia. It is one of the bigger and brighter near-Earth asteroids that have been discovered and it has recently been found to have two moons (Benner et al., 2017). It was observed with the Harvard Clay Telescope in the R-band over the course of seven nights, though three were unusable due to poor weather and high scatter. We found the rotational period to be $2.3580 \pm 0.0015$ h, which agrees well with the previously determined rotational period of $2.3580 \pm 0.0002$ h (Warner, 2016).

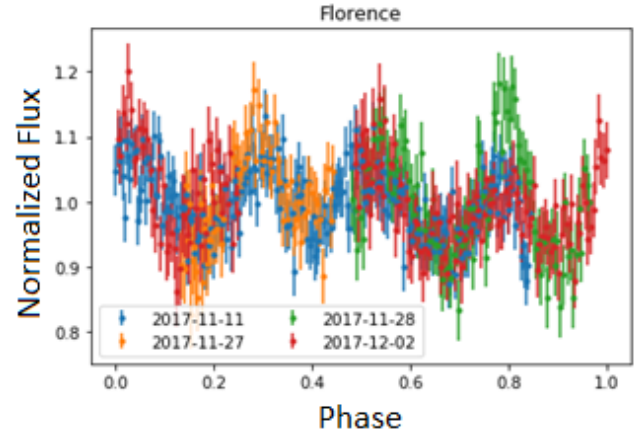

3830 Trelleborg was discovered 1986 September 11 at the Brofelde Observatory in Denmark. It was observed with KeplerCam in the i-band over the course of nine nights. Since we observed it over a long period of time, the visual magnitude of the asteroid changed over the course of the observations, so we did absolute photometry using *AstroImageJ* and normalized the lightcurve by subtracting the magnitudes corresponding to a linear brightness decline. We then determined the rotational period as described above and found the following phase diagram with a period of $17.059 \pm 0.017$ h.

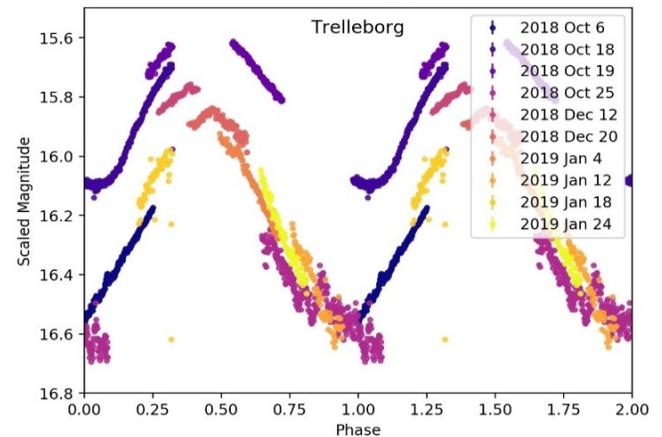

| Number | Name | 20yy/mm/dd | Phase | $L_{PAB}$ | $B_{PAB}$ | Period (h) | P.E. | Amp | A.E. | Grp |
|---|---|---|---|---|---|---|---|---|---|---|
| 3122 | Florence | 17/10/02-12/01 | 77.9,20.7 | 55 | 51 | 2.3580 | 0.0015 | 0.17 | 0.04 | Amor |
| 3830 | Trelleborg | *18/10/06-01/24 | 6.7,18.9 | 356 | 9 | 17.059 | 0.017 | 0.76 | 0.01 | Eos |
| 131077 | 2000 YH105 | 19/01/17-02/08 | 8.9,13.7 | 121 | 12 | 1.8130 | 0.0003 | 0.19 | 0.02 | MB-O |

Table I. Observing circumstances and results. *Observations extended into 2019. The phase angle is given for the first and last date. LPAB and BPAB are the approximate phase angle bisector longitude and latitude at mid-date range (see Harris et al., 1984). Grp is the asteroid family/group (Warner et al., 2009). MB-O: outer main-belt.

Due to intrinsic variability of the asteroid, there were variations in the peak heights, so we manually shifted the magnitude. This had no effect on the measured rotational period.

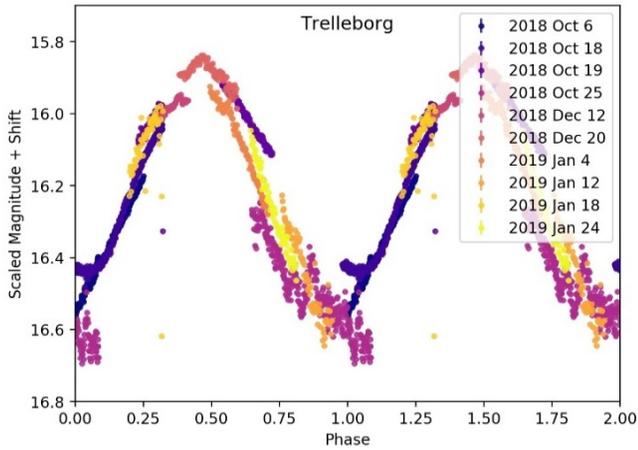

(131077) 2000 YH105 was discovered 2000 December 28 in Socorro, New Mexico. We observed it over the course of seven nights with KeplerCam in the i-band. We were the first to measure its rotational period and found that it is $1.813 \pm 0.00003$ h. We realize this period is short and would be rare for an asteroid this diameter, so we performed a second analysis on the data. Using the Phase Dispersion Minimization (PDM) technique (Plavchan et al. 2008), we found the same period again. We cannot completely rule out a multiple of our best period being the true period given our current data. Ultimately more data is required for confirmation of the suggested period.

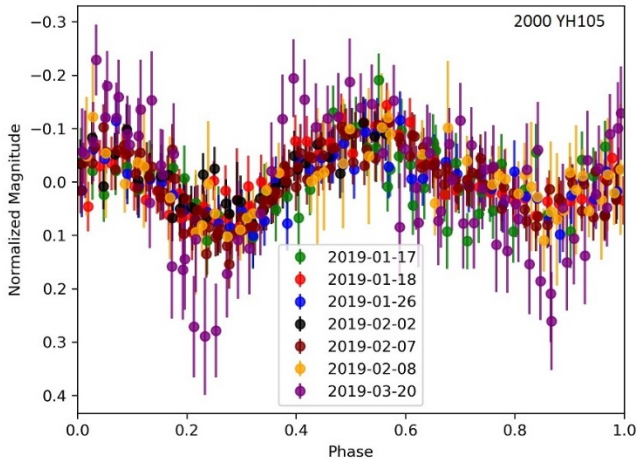